\documentclass[]{aa} 
 
\usepackage{natbib}  

\usepackage{graphicx}
\usepackage[varg]{txfonts}
\usepackage{hyperref}

\def\Ou  {\ion{O}{i}}

\def\Nau  {\ion{Na}{i}}
\def\Cau  {\ion{Ca}{i}} 
\def\Cad  {\ion{Ca}{ii}} 
\def\Feu  {\ion{Fe}{i}}
\def\Fed  {\ion{Fe}{ii}}
\def\Tiu  {\ion{Ti}{i}}

\def\Mgu  {\ion{Mg}{i}}

\def\Niu  {\ion{Ni}{i}}
\def\Srd  {\ion{Sr}{ii}} 
\def\Bad  {\ion{Ba}{ii}}

\def\Cdou {$\rm ^{12}C$}
\def\Ctre {$\rm ^{13}C$}
\def\Cd    {$\rm C_{2}$}
\def\Od    {$\rm O_{2}$}
\def\hm    {$\rm H^{-}$}
\def\HdO   {$\rm H_{2}O$}
\def\Teff  {$T_\mathrm{eff}$}
\def\logg  {$\log g$}
\def\loggf {$\log gf$}
\def\vt    {$\rm v_{t}$}
\def\kms   {$\rm km\,s^{-1}$}
\def\ms    {$\rm  m\,s^{-1}$}
\def\kiex  {$\rm \chi_{ex}$}
\def\Pis   {Pisces\,II\,10694}

\begin{document}

\title{
A CEMP-no star in the ultra-faint dwarf galaxy Pisces II
\thanks{Based on observations collected at the European Organisation for Astronomical Research in the Southern Hemisphere under ESO programme  099.B-0062(A).}
}
\author {
Spite M.\inst{1}\and 
Spite F.\inst{1}\and
Fran\c cois P.\inst{1,2}\and
Bonifacio P. \inst{1}\and
Caffau E.\inst{1}\and
Salvadori S. \inst{1,3,4}
}

\institute {
Observatoire de Paris, PSL University, CNRS, GEPI,
Place Jules Janssen, 92195 Meudon, France
\and
Universit\'e de Picardie Jules Verne, 33 rue St-Leu, 80080, Amiens, France
\and
Dipartimento di Fisica e Astronomia, Universit\'
a di Firenze, Via G. Sansone 1, Sesto Fiorentino, Italy
\and
INAF/Osservatorio Astrofisico di Arcetri, Largo E. Fermi 5, Firenze, Italy
}


\authorrunning{Spite et al.}

\titlerunning{Light neutron-capture elements in EMP stars}

  \abstract
{}
{A probable carbon enhanced metal-poor  (CEMP) star, \Pis, was discovered recently in the ultra-faint (UFD) galaxy Pisces II. This galaxy is supposed to be very old, suspected to include dark matter,
 and  likely formed the bulk of its stars before the reionisation of the Universe.
}
{
New abundances have been obtained from observations of \Pis~ at the Kueyen ESO VLT telescope, using the high-efficiency spectrograph: X-Shooter.
}
{
We found that \Pis~ is a CEMP-no star with [Fe/H]=--2.60\,dex. Careful measurements of the CH and \Cd~ bands confirm the enhancement of the C abundance ([C/Fe]=+1.23). This cool giant has very probably undergone  extra mixing and thus its original C abundance could be even higher. Nitrogen, O, Na, and Mg are also strongly enhanced,  but  from Ca to Ni the ratios [X/Fe] are similar to those observed in classical very metal-poor stars.
With its low Ba abundance ([Ba/Fe] =--1.10 dex) \Pis~ is a CEMP-no star. No variation in the radial velocity could be detected between 2015 and 2017.
The pattern of the elements has a shape similar to the pattern found in galactic CEMP-no stars like CS 22949-037 ([Fe/H]=--4.0) or SDSS J1349+1407 ([Fe/H]=--3.6).
}
{The existence of a CEMP-no star in the UFD galaxy Pisc\,II suggests that this small galaxy likely hosted zero-metallicity stars. This is consistent with theoretical predictions of cosmological models supporting the idea that UFD galaxies are the living fossils  of the first star-forming systems.
}
\maketitle

%

The ultra-faint dwarf (UFD) galaxies contain the largest fraction of metal-poor stars of any galaxy type \citep[e.g.][]{KirbySG08} and they could be the direct descendants of the first generation of galaxies in the Universe \citep{BovillRico09,SalvadoriFer09}. It has been shown that the chemical composition of the UFD stars  is very similar to the chemical composition of the extremely metal-poor (EMP) galactic stars, \citep[see e.g. ][]{FrancoisMB16,JiFS16,JiFE16}.

In our Galaxy, in the `normal' EMP turn-off stars, the abundance ratios of the elements with respect to Fe is about the same as that found in  solar-type stars, except for the $\alpha$ elements which are slightly enhanced with respect to iron (e.g. $\rm[Ca/Fe] \approx 0.3\,dex$). In these stars, C is also slightly enhanced: [C/Fe] has been found equal to $\rm 0.45 \pm 0.10$ \citep{BonifacioSC09}. Many giants undergo a deep mixing that transforms the atmospheric carbon into nitrogen, and consequently their carbon abundance is systematically lowered relative to turn-off stars,  sometimes with [C/Fe]<0.  The abundance of the elements heavier than Zn (neutron-capture elements) is very scattered in the classical EMP turn-off and giant stars \citep{FrancoisDH07,SpiteSB18}. 

At low metallicity many stars are carbon-enhanced compared to the normal EMP stars. If we sample stars of lower and lower metallicity, we find larger and larger fractions of carbon-enhanced stars. About 30\%\ to 40\% of the EMP stars with $\rm[Fe/H] \approx -3$ possess significant overabundances of carbon relative to iron. This fraction rises to at least 80\% for stars with [Fe/H] <--4.0  \citep{TimBeers18}, and  among the twelve known stars with 60 000 times less metals than the Sun ($\rm[Fe/H]<-4.5$), only one star \citep{CaffauBF11,BonifacioCS15} has a confirmed carbon abundance not compatible with the definition of the CEMP stars  adopted here: $\rm[C/Fe]>+1.0$ \citep{BeersC05}.

The carbon-enhanced metal-poor (CEMP) stars have been extensively studied in our Galaxy at low and high resolution 
\citep[e.g.][ and references therein]{AllenRR12,LeeBM13,LeeSB14,HansenNH16,YoonBP16}.
They are characterised by a strong overabundance of carbon, and very often an overabundance of nitrogen, oxygen, and magnesium. Moreover, they have very diverse abundances of the neutron-capture elements: some are very strongly polluted by neutron-capture elements built by the slow (`s')  and sometimes the rapid (`r') processes (here we  call all of them CEMP-s). Others CEMP stars do not show any enhancement of the neutron-capture elements compared to the normal metal-poor stars and they are called `CEMP-no'.
In our Galaxy, all the most metal-poor carbon-rich stars ($\rm[Fe/H] < -3.6$)  belong, as far as we know, to the group of the CEMP-no  \citep{SpiteCB13,BonifacioCS15,CaffauGB18}: their carbon abondance A(C) is always between A(C)=5.5 and 7.7. At higher metallicity some stars are CEMP-no, but most of them are CEMP-s: their carbon abundance  can reach A(C)=8.7, their Ba abundance is very high, and they have generally been found to be binary stars \citep[see e.g.][]{CaffauGB18}. According to several authors \citep[e.g.][]{MasseronJP10,AbatePS15}, their chemical pattern suggests a previous mass-transfer from their companion in its asymptotic giant branch (AGB) phase.

Since  their discovery it has been proposed that  the CEMP-no stars  formed in an environment polluted by the first stellar generations \citep[e.g.][]{BonifacioLC03,NomotoUM03}. This idea has been further confirmed by several authors who suggest that the carbon in the CEMP-no stars was  brought by faint supernovae of zero metallicity \citep{CookeMadau14,deBennassutiSV14,BonifacioCS15}. Probably the C-enhancement made it easier for them to  form from a gas depleted in other metals, and the large scatter of the carbon abundance displayed in these stars (about 1 dex) is the result of the nucleosynthesis of a limited number of faint supernovae of zero metallicity that have polluted the gas. 
If this scenario is correct, then many CEMP-no stars should be also found in dwarf galaxies, and especially in UFD galaxies, which are the faintest ($\rm L < 10^{5}L_{\odot}$), the more metal-poor, and likely the oldest galactic systems \citep{SalvadoriST15}.
In  recent years, many metal-poor stars have been observed and analysed in the dwarf galaxies of the Local Group,  and several teams have been looking for carbon-rich stars \citep[e.g.][]{NorrisGW10,LaiLB11,FrebelSK14,FrebelNG16,SkuladottirTS15,ChitiSF18}. To date, however,  only five true CEMP stars ($\rm[Fe/H] \leq -2.5 ~and~ [C/Fe] > +1$) have been observed in these faint systems. All these stars are CEMP-no.

Recently \citet{KirbySC15} have observed spectroscopically seven stars in the UFD galaxy Pisces II ($\rm L  = 10^{4}L_{\odot}$), discovered by \citet{BelokurovWE10};  however,  he was able to  measure the metallicity of only four of them. Among these four stars, one, \Pis, is a CEMP star with [Fe/H]= --2.7. In the Ultra Faint Dwarfs very few stars are observable. Establishing that even one of the stars is a CEMP-no star (or a CEMP-s) can provide useful information. The quality of the spectra of \citet{KirbySC15} did not allow him  to determine the C abundance (although the C-enhancement is visible from the CN band between 830 nm and 840 nm) or the abundance of any element other than iron. This star is a cool giant, and thus the probability that it has undergone  deep mixing is high. In this case the carbon abundance in the gas that formed the star should have been even  higher than observed now in the atmosphere of the star, the abundance of the heavier elements remaining untouched by deep mixing.

The aim of this paper is to determine the abundances of the main elements in \Pis~ from spectra obtained with a higher resolution and in a larger wavelength range than the spectrum studied by \citet{KirbySC15} in order to establish  the CEMP type of this star.

\begin{figure}
\resizebox{\hsize}{!}                   
{\includegraphics[clip=true]{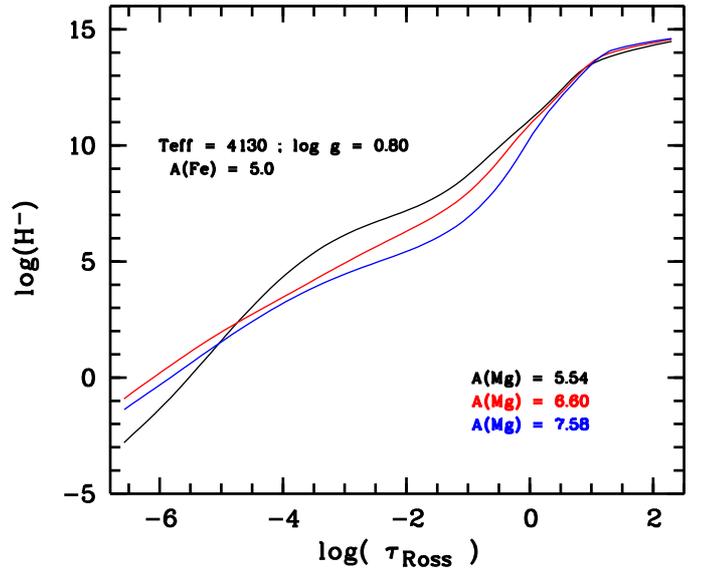} }
\caption[]{Number density of \hm~vs. optical depth $\rm log\,\tau_{Ross}$ for three different abundances of Mg. The blue line (A(Mg) = 7.58) corresponds to the solar abundance of Mg, the red line to the Mg abundance in \Pis, and the black line (A(Mg) = 5.54) to the Mg abundance expected for a normal metal-poor star with [Fe/H]=--2.6. The molecular bands and the metallic lines are mainly formed  between $\rm log\,\tau_{Ross} = -2$ and --1, a region where the number density of \hm\ is strongly affected by the Mg abundance.
}
\label {mghm}
\end{figure}

\begin{figure*}[ht]
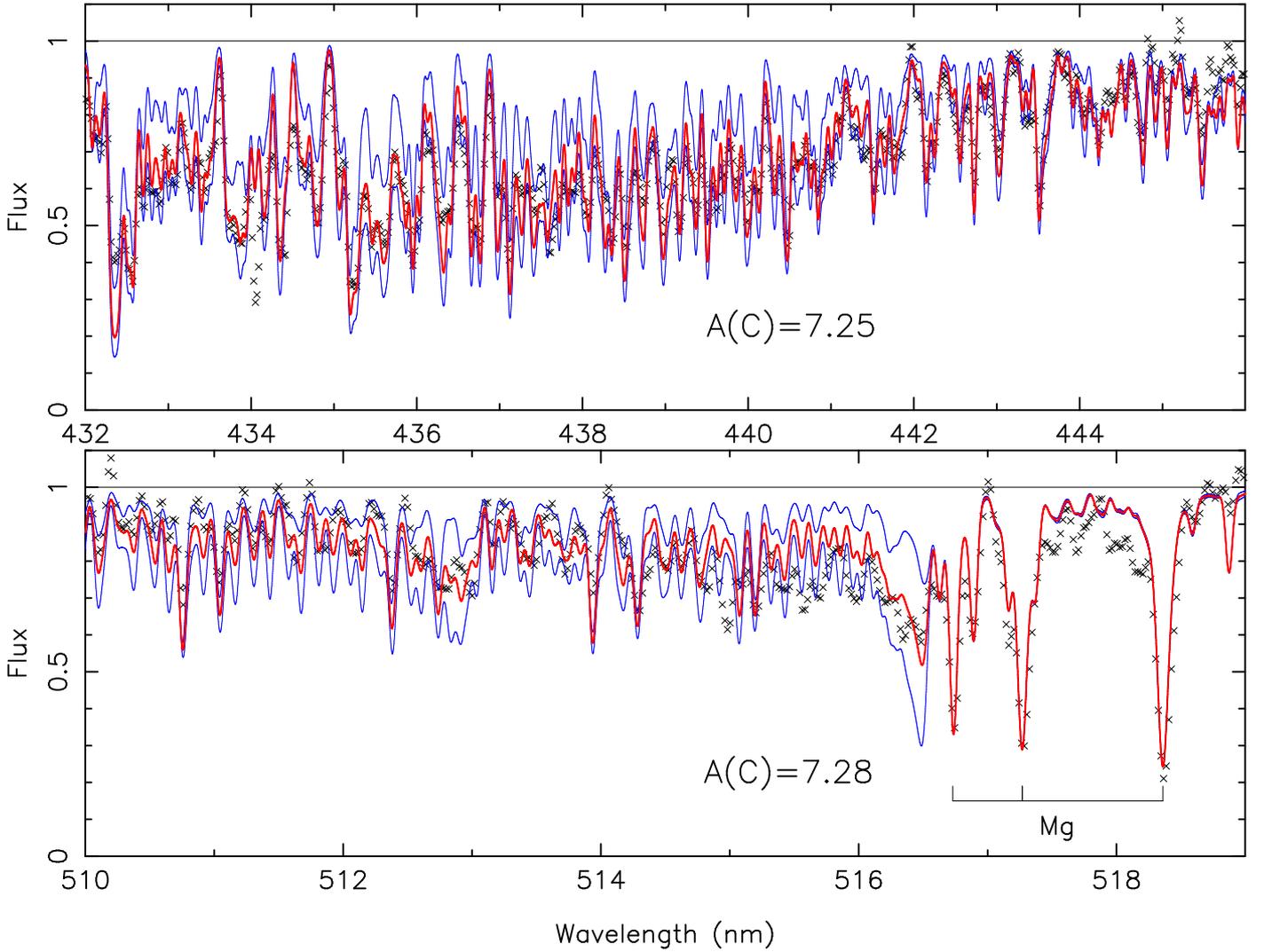

\resizebox{\hsize}{!}                   
{\includegraphics {PisCH.ps}}
\resizebox{\hsize}{!}
{\includegraphics {PisC2swan.ps}}
\caption[]{Observed spectrum (black crosses) and synthetic profiles of the CH and \Cd~ bands 
computed with A(C)=6.8 and 7.4 for the CH band, and A(C)=7.1 and A(C)=7.4 for the \Cd~ band (blue lines). The red line represents the synthetic spectrum computed with the adopted value of the C abundance in this region of the spectrum (A(C)=7.28 for the \Cd band and 7.25 for this region of the CH band; see Table \ref{lines}). The abscissa is the wavelength in nm. Shown on the red side of the \Cd~band is the good fit of the Mg triplet obtained with A(Mg)=6.6
}
\label {CHC2}
\end{figure*}

\section {Observational data} 
The {\sl g} magnitude of \Pis~is 19.9. This very faint star was observed in service mode with Kueyen (ESO-VLT, UT2) and the high-efficiency spectrograph X-Shooter \citep{DOdoricoDM06,VernetDD11}.  The observations were performed in staring mode with 1x1 binning and the integral field unit \citep{GuinouardHP-IFU06}.  The stellar light is divided in three arms by X-Shooter; we analysed here only the UVB spectra (from 400 to 560\,nm with a resolving power $R = 7900$) and the visual (VIS) spectra (from 560 to 1000\,nm with $R = 12600$). The signal-to-noise ratio (S/N) of the infrared spectrum ($\rm\lambda > 1000\,nm$) is too low to allow the analysis. The spectra were reduced using the same procedure as in \citet{CaffauBF13}. 
The total exposure time was 11 hours divided in thirteen 3000\,s spectra.

\subsection{Radial velocities measurements}

Thirteen spectra were obtained in 2017 between June 30 and August 20. The radial velocity has only been measured on  the twelve spectra with the best S/N ratio (Table \ref{radvel}). The geocentric radial velocity has been established from the position of the $\rm H_{\alpha}$ line, which is well defined in all the spectra.\\
Since X-shooter is a single object spectrograph for the Cassegrain focus, it is sensitive to flexures.  The flexure correction can be determined from the apparent position of the telluric lines (zero point); for precise measurement of the radial velocity it is very important to take into account this correction.
\begin{table}
\caption[]{ 
Radial velocity of \Pis~ measured on the twelve best spectra. All the radial velocities are given in \kms.
 The barycentric radial velocity is equal to (geocentric Rad. Vel.) -- (telluric Rad. Vel.) + (barycentric correction).}
\label{radvel}
\begin{tabular}{l@{~~}c@{~~}c@{~~}c@{~~}c@{~~}c@{~~}c@{~~}c@{~~}c@{~~}c@{~~}c@{~~}c@{~~}c}
\hline
    &geocentric& telluric lines& barycentric& barycentric\\ 
MJD & Rad. Vel.&  Rad. Vel.    & correction & Rad. Vel.  \\
\hline
57934.35 & -268.4  & -4.4   & 26.9 & -237.1 \\
57955.28 & -261.0  & -2.2   & 21.9 & -236.9 \\
57955.33 & -255.4  & -3.4   & 21.7 & -230.3 \\
57956.26 & -261.7  & -4.0   & 21.7 & -236.0 \\
57963.22 & -251.6  &  0.6   & 19.3 & -232.9 \\
57964.22 & -249.7  &  0.9   & 18.9 & -231.7 \\
57964.31 & -252.0  &  0.8   & 18.7 & -234.1 \\
57965.26 & -251.1  &  3.0   & 18.5 & -235.6 \\
57980.17 & -264.2  &-15.9   & 12.5 & -235.8 \\
57982.20 & -258.2  &-13.0   & 11.5 & -233.6 \\
57985.24 & -265.1  &-16.9   &  9.9 & -238.3 \\
57985.29 & -258.8  &-15.9   &  9.8 & -233.1 \\
\hline    
\end{tabular}  
\end{table}

The uncertainty in the individual measurement of the barycentric radial velocity is less than 4.0\,\kms. 
From the measurements listed in Table \ref{radvel} we were  not able to detect a variation in the radial velocity between June and August 2017.  The mean barycentric radial velocity is $-234.6 \pm 2.4$ \kms, which is consistent \footnote {For each star in Pisces,  Kirby gives the radial velocity relative to the centre of the Sun (heliocentric radial velocity) neglecting the rotation of the Sun around the gravity centre of the solar system, but the difference between the heliocentric and barycentric velocities is of the order of a few \ms, which is negligible compared to the measurement errors.}  with the value measured on May 2015 \citep{KirbySC15}: $-232.0 \pm 1.6$ \kms.

\section {Abundance analysis}
In a first step we interpolated a model in a grid of LTE OSMARCS  models \citep{GustafssonBE75,GustafssonEE03,GustafssonEE08} with the atmospheric parameters given by \citet{KirbySC15}: \Teff=4130K, \logg=0.8. These parameters are mainly based on photometry and theoretical isochrones from the Yonsei-Yale group \citep{DemarqueWK04}.
With this model we found in \Pis~ a very strong enrichment of C, N, and Mg relative to Fe; our spectrum does not contain \Fed~ lines, which could constrain  the gravity better.

In these cool metal-poor giants, the main source of opacity is the negative hydrogen ion \hm~ whose abundance depends on the electron density. The main electron donor is always Mg,  especially in the atmosphere of \Pis~ since all the other metals are very scarce. As a consequence the continuous absorption strongly depends on the abundance of magnesium. In Fig. \ref{mghm} we show the influence of the magnesium abundance on the number density of \hm.  The metallic lines are formed at an optical depth close to $\rm log\,\tau_{Ross} = -1$, and the molecular bands nearer the surface, close to or below $\rm log\,\tau_{Ross} = -2$, regions where the number density of \hm~ is strongly impacted by the magnesium abundance. It is thus particularly important to determine the Mg abundance before computing the  molecular lines of \Cd~ and CN.

As a consequence, in a second step we computed an ATLAS12 model  \citep{Kurucz05} with these peculiar  overabundances of C, N, and Mg, and we iterated to determine the abundances of the different elements. We note that after the NLTE correction, there is a good agreement between the Ca abundance deduced from the \Cau~ and the \Cad~ lines (Table \ref{abund}) confirming the choice of the gravity.
The parameters of the model we finally used for the analysis of \Pis~ are: \Teff=4130\,K, \logg=0.8, \vt=2.0\,\kms, [M/H]=--2.5, A(C)=7.0, A(N)=7.0 and A(Mg)=6.6.\\
The abundance analysis was performed using the LTE spectral line analysis code {\tt turbospectrum} \citep{AlvarezP98,Plez12}.
Since the resolution of the spectra is not very high and since there are many molecular lines all along the spectrum, the lines of the atomic elements are often blended. As a consequence we determined the abundances by fitting synthetic spectra to all visible lines.
The results of these computations and the main characteristics of these lines are given in the Appendix (Table \ref{lines}). The hyperfine structure has been taken into account,  in particular for the \Bad~ lines.\\
The abundance of sodium is deduced from the Na D lines (Table \ref{lines}) which are known to present a very strong non-LTE effect in cool metal-poor giants \citep{AndrievskySK07}. However, following \citet{MashonkinaSS00},  at [Fe/H]=--2 this effect is maximum around \Teff=5000K and decreases very rapidly when the temperature decreases. At  $\rm T_{eff} \approx 4100$K and  \logg $\approx 1.0$, (as in \Pis), the NLTE correction is negligible.

The mean abundances are given in Table \ref{abund}. 
Since the red \Cad~lines present a very strong non-LTE effect in cool giants, we adopted in Table \ref{abund} a correction of --0.3\,dex for the red \Cad~ triplet (based on \citealt{SpiteAS12}).\\

\begin{table}
\begin{center}   
\caption[]{ 
Mean abundance of the elements in the Sun, in \Pis, the number of lines used for this determination, standard error, $\rm [X/H]= A(X)_{\star}-A(X)_{\odot}$, and 
$\rm [X/Fe]= [X/H]_{\star}-[Fe/H]_{\star}$. 
The solar abundances are from \citet{LoddersPG09} for all elements except Fe and C, which are from \citet{CaffauBF11}.
  }
\label{abund}
\begin{tabular}{clcrcrrrrrrrrrrr}
\hline
Element & $\rm A(X)_{\odot}$& $\rm A(X)_{\star}$& N & err & [X/H] &  [X/Fe] \\
\hline
Li     &       &$\le0.2$& \\
C (CH) &  8.50 &   6.98 & - & -   & --1.52 & +1.08\\
C (\Cd)&  8.50 &   7.28 & - & -   & --1.23 & +1.38\\
N (CN) &  7.86 &   7.09 & - & -   & --0.77 & +1.81\\
$\rm [\Ou]$&8.76&  8.43 & 2 & 0.18& --0.33 & +2.25\\
\Nau   &  6.30 &   4.73 & 1 &  -  & --1.57 & +1.03\\
\Mgu   &  7.54 &   6.56 & 7 & 0.28& --0.98 & +1.60\\
\Cau   &  6.33 &   3.88 & 7 & 0.23& --2.45 & +0.13\\
\Cad*  &  6.33 &   3.94 & 3 & 0.19& --2.37 & +0.21\\
\Tiu   &  4.90 &   2.74 & 1 & -   & --2.16 & +0.42\\
\Feu   &  7.52 &   4.92 &15 & 0.10& --2.60 & -~~~~\\
\Niu   &  6.23 &   3.83 & 2 & 0.23& --2.40 & +0.20\\
\Srd   &  2.92 &  0.5: & 2 & -   & --2.42:&  +0.18:\\
\Bad   &  2.18 &  -1.52 & 3 & 0.18& --3.70 & --1.10\\
\hline    
\multicolumn{7}{l}{* abundance corrected from non-LTE effects (see text).}
\end{tabular}  
\end{center}   
\end{table}

\subsection{{\bf Abundance of the light elements: Li, C, N, and O}}   \label{C-N}
We tried to measure Li  in \Pis.  A $\rm\chi$-squared fitting gives $\rm A(Li) = 0.0\,dex$, but in a conservative way we estimated that $\rm A(Li)\le +0.2\,dex$.

The abundance of C and N is deduced from the molecular bands of CH, \Cd, and CN. The data of these molecular lines are taken from \citet{Plez08} and \citet{Plez18}. 
In Fig. \ref{CHC2}, as an example, we show the fit of a part of the CH and the \Cd~ bands. We adopted for the C abundance the mean of the abundances deduced from the CH and the \Cd~ lines: A(C)=7.13 (see Tables \ref{abund} and \ref{lines}). The resolution of the spectrum is not sufficient to determine the ratio  \Cdou/\Ctre.

\begin{figure}
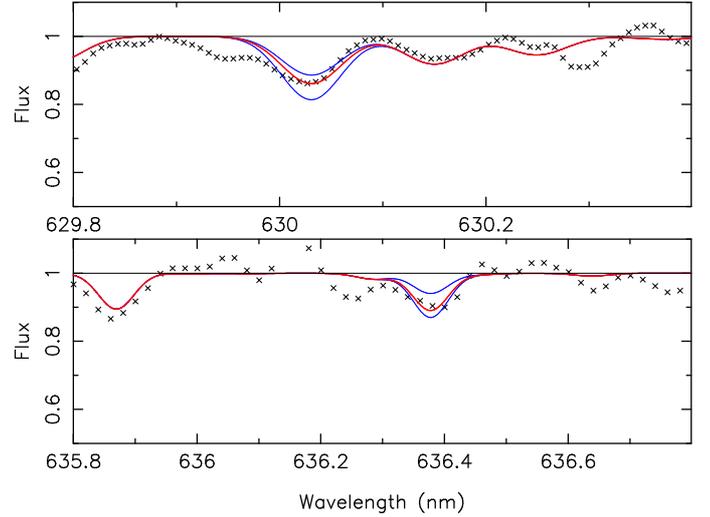

\centering
\resizebox{\hsize}{!}
{\includegraphics {PisOxy6300.ps} }
\resizebox{\hsize}{!}
{\includegraphics {PisOxy6363.ps} }
\caption[]{Fit of the profile of the forbidden oxygen lines. The abscissa is the wavelength in nm. The blue lines show the synthetic spectra computed with A(O)=8.1 and A(O)=8.7. The red line is the best fit obtained with A(O)=8.3 for the line at 630\,nm, and A(O)=8.55 for the line at 636.3\,nm.} 
\label{oxy}
\end{figure}

The forbidden oxygen lines at 630.03 and 636.38 nm are blended by telluric bands of \Od~ and \HdO. The lines of these bands are always weak in our spectra; however, we accounted for them by dividing each of the observed spectra by the corresponding spectrum  of the telluric lines \footnote{http://cds-espri.ipsl.fr/tapas/} \citep{BertauxLF12,BertauxLF14} computed at the time of the observation and at the position of the star. The individual spectra free from telluric lines are then added. The results of these computations are shown in Fig. \ref{oxy}.

\begin{figure}[ht]
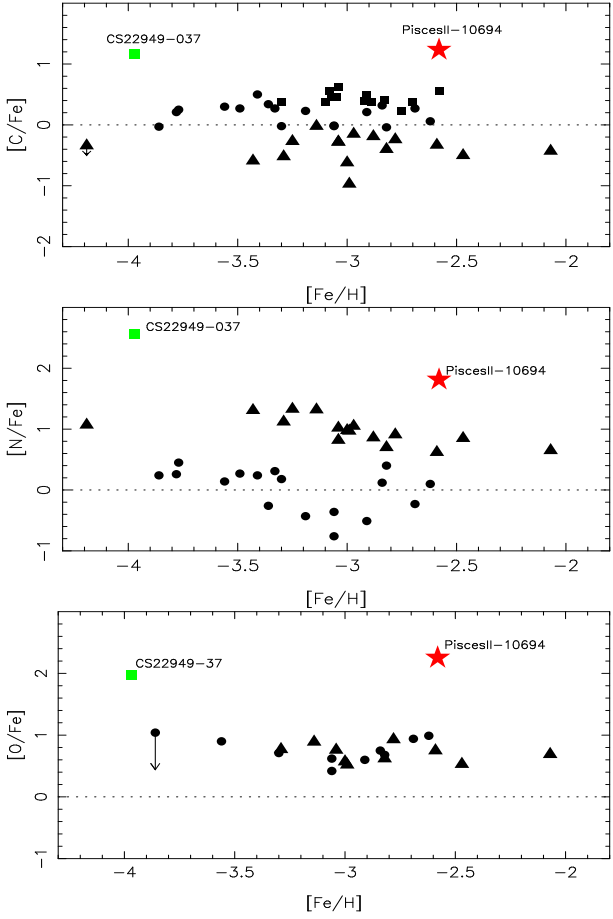

\centering
\resizebox  {8.0cm}{4.0cm}
{\includegraphics {abgfe-cmix.ps} }
\resizebox  {8.0cm}{4.0cm}
{\includegraphics {abgfe-nmix.ps} }
\resizebox  {8.0cm}{4.0cm}
{\includegraphics {abgfe-oamix.ps} }
\caption[]{[C/Fe], [N/Fe], and [O/Fe] vs. [Fe/H] for \Pis~ and a sample of normal galactic very metal-poor stars. 
The dwarfs are represented by solid squares, the stars in the lower RGB branch by solid circles, and the mixed stars by triangles. 
The classical galactic CEMP-no giant  CS\,22949-037 is added (green square). The star \Pis~ is represented by a red star. The large overabundances of C, N, and O in \Pis~ are very similar to those observed in CS\,22949-37.
} 
\label{CNO-LP}
\end{figure}

In Fig. \ref{CNO-LP} we compare the C, N, and O abundances in \Pis~and in carbon-normal metal-poor stars studied homogeneously in the frame of the ESO Large Programme `First Stars' \citep{SpiteCP05,SpiteCH06,BonifacioSC09}.
We added also a classical carbon-rich very metal-poor giant studied in the frame of this Large Programme: CS\,2249-37 \citep{DepagneHS02}. Mixing processes explain the large scatter observed in the C and N abundances of the carbon-normal metal-poor stars. In  giant stars, at different levels of the evolution,  mixing occurs between the deep layers and the atmosphere, bringing to the surface CNO processed material: the C abundance decreases and the N abundance increases, but [O/Fe] remains constant.  From their atmospheric parameters, \Pis~ and CS\,22949-37 are both mixed giants and should be compared to the mixed giants in Fig. \ref{CNO-LP} (black triangles). Both stars appear to be very enhanced in C, N, and O, and the enhancements are similar.

\begin{figure}
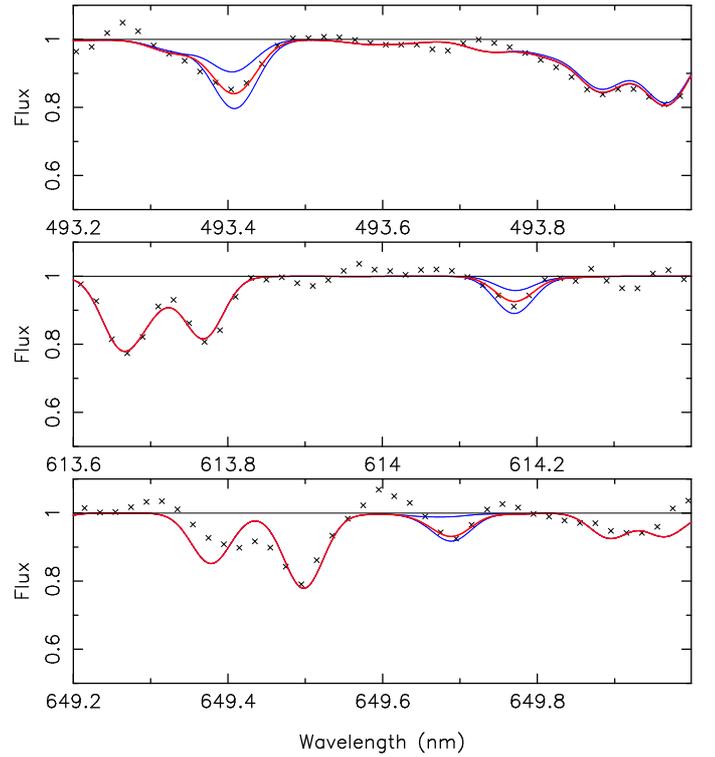

\centering
\resizebox{\hsize}{!}
{\includegraphics {PisBa4934.ps} }
\resizebox{\hsize}{!}
{\includegraphics {PisBa6141.ps} }
\resizebox{\hsize}{!}
{\includegraphics {PisBa6496.ps} }
\caption[]{Fit of the profile of the Ba lines. The abscissa is the wavelength in nm. The blue lines show the synthetic spectrum computed with A(Ba)=--2.4 and A(Ba)=--1.2. The red line is the best fit obtained with A(Ba)=--1.66 for the line at 493.4\,nm, A(Ba)=--1.58 for the line at 614.1\,nm, and --1.31 for the line at 649.6\,nm.} 
\label{Ba}
\end{figure}

\subsection{Abundance of the neutron-capture elements}
On the spectra of \Pis~ we could only measure the Sr and Ba abundances. Only two Sr lines are visible: one (at 407.76\,nm) is very blended in a region where the S/N of the UVB spectrum is low; the other line (at 1003.66\,nm) is located at the very end of the VIS spectrum. In both cases the measurement of the abundance is very uncertain and we have estimated that the error of [Sr/Fe] is close to 0.5\,dex.\\
We were able to  measure four Ba lines. The resonance line of \Bad~ (at 455.4 nm) is in a very crowded region of the spectrum with a rather low S/N ratio; its measurement is very uncertain, and thus we have determined the abundance of Ba {only} from the other three  lines (Table \ref{lines} and Fig. \ref{Ba}).

\begin{figure}
\centering
\resizebox {\hsize}{!}
{\includegraphics {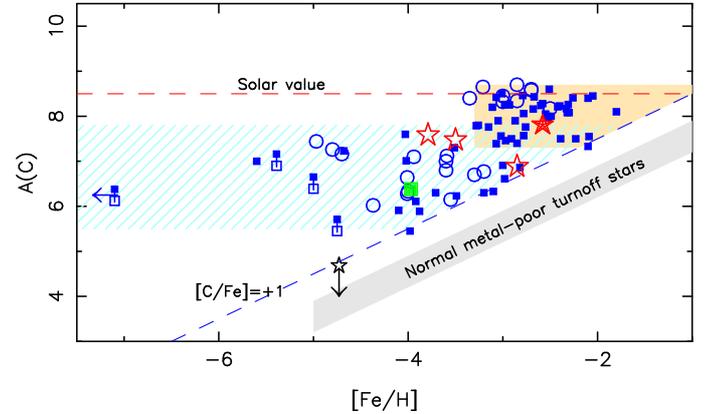} }
\caption[]{Abundance of carbon A(C) vs. [Fe/H] in dwarfs and turn-off galactic CEMP stars from the literature (blue squares). Only the determinations obtained from high-resolution spectra were selected. 
The measurements made homogeneously by our team are represented with open blue circles.
The dashed blue line (representing $\rm[C/Fe]=+1$) separates the region of the carbon-rich metal-poor stars from the region of the normal metal-poor stars. On the upper A(C) band (yellow zone) all the stars but four are Ba-rich. On the contrary, on the lower A(C) band (hatched blue zone) all the stars are CEMP-no. Below [Fe/H]=--3.4 all the CEMP stars belong to the second group.\\ The \Pis~ star is represented by a red filled star  and the CEMP giants in the dwarf galaxies Sculptor, Boo\,I and Seg\,I, by open red stars. The galactic giant CS\,22949-037 is represented by a filled green square.
} 
\label{Cab}
\end{figure}

\section{Discussion}

\citet{KirbySC15} consider that \Pis~ belongs to the UFD galaxy Pisces II, mainly from the radial velocity of the star. With the Gaia Data Release 2 (DR2)  \citep{Gaia-Prusti16,Gaia-Brown18,ArenouLB18},  it is possible in some Local Group galaxies to use the proper motions and even the parallaxes of the stars to confirm that they are true members \citep[e.g.][]{Gaia-Helmi18,Simon18,FritzBP18}. 
At present this is not possible for Pisces\,II because these quantities are now available for only two spectroscopic members of Pisces II (9004 and 10694). They have been used to infer a systemic proper motion for the system \citep{FritzBP18}, but the  large uncertainty of the parallaxes and proper motions hamper the possibility of safely identifying them as possible foreground stars.

\subsection{The CEMP type of \Pis}

\subsubsection{Abundance of C, N, O} \label{mixing}

In Fig. \ref{Cab} we have plotted the abundance of C versus [Fe/H] for galactic metal-poor {turn-off} stars \citep{SpiteCB13,BonifacioCS15}. In turn-off stars, the original C abundance is not diluted by mixing with deep layers where C is transformed into N, and thus the C abundance measured in these stars is expected to be the abundance of carbon in the cloud from which the star formed. 
The `normal' metal-poor turn-off stars  like those studied in \citet{BonifacioCS15} in the frame of the ESO Large Programme `First Stars' from high-resolution, high S/N spectra are located in the grey zone of the figure. Their relatively high C abundance (measured from the CH band): ($\rm[C/Fe]=+0.45 \pm 0.10$) is in agreement with the models of the chemical evolution of the Galaxy \citep[see e.g.][]{RomanoKT10}.  
We thus adopted  $\rm[C/Fe]\ge +1$ as a definition of the C-rich stars to avoid the tail of the distribution of the normal metal-poor stars as already proposed by \citet{BeersC05}.

With this definition of the C-rich stars, the carbon-rich turn-off stars, in  Fig. \ref{Cab} are located inside two different bands \citep{SpiteCB13,BonifacioCS15}.\\ 
Most of the stars in the highest carbon band are Ba-rich and sometimes also Pb-rich; however, some rare stars have been found to be CEMP-no.\\
On the contrary, all the stars in the lowest hatched blue C band  have been found to be  CEMP-no \citep [see  Fig. 6 in][for more details]{BonifacioCS15}. Below [Fe/H]=--3.4  all the CEMP stars belong to this second group.\\
It is generally considered that the CEMP-no stars were born with the observed chemical composition. On the contrary, the chemical composition of the  CEMP-s  stars is explained by a mass-transfer from a companion in its asymptotic giant branch (AGB) phase. Most of the CEMP-s stars show  radial velocity variations, which suggests that they are all binaries \citep{LucatelloTB05,StarkenburgSS14}.\\

If we want to compare the C, N, O abundances in stars of our Galaxy and in the carbon-rich \Pis, we have to take into account that it is not a turn-off star and that  extra mixing could occur, inducing a transformation of C into N and thus a decrease in the C abundance. In  Fig. \ref{CNO-LP} (upper panel) the difference in [C/Fe] between turn-off stars and evolved mixed giants is about +0.7 dex (see also  \citealt{SpiteCP05,SpiteCH06,BonifacioSC09}). In the same way, in Fig. \ref{CNO-LP} (middle panel) we estimate that the correction for [N/Fe] is about --0.8dex.\\ 

In Fig. \ref{Cab} we have added the position of \Pis~ (filled red star symbol) and as a comparison, the position of the classical galactic evolved C-rich giant CS\,22949-037 \citet{DepagneHS02}.
Both stars are cool giants with \logg=0.8 and 1.5\,dex and their C abundance has to be corrected for extra mixing. We adopted the same correction for these two stars (+0.7 dex).

Two CEMP stars have been identified in the dwarf galaxy  Sculptor \citep{SkuladottirTS15,ChitiSF18}, but following our definition of the C-rich stars, only one can be considered  a CEMP: Sculptor\,11-1-4422 with [C/Fe]=1.26\,dex, after mixing correction. In the UFD galaxy Bootes\,I, \citet{LaiLB11} measured for Boo21 [C/Fe]=+2.20, or [C/Fe]=+2.90 after mixing correction. This star is also known as  Boo-119 \citep{GilmoreNM13,FrebelNG16}. In the UFD galaxy Segue\,I, a CEMP star has been also detected: Segue\,I-7 with [C/Fe]=+2.30, or [C/Fe]=+2.50 after mixing correction \citep{NorrisWG10}.
These stars have been also plotted in Fig. \ref{Cab}.
The stars Sculptor\,11-1-4422, Boo\,I-119, and Segue\,I-7   clearly belong to the lower A(C) band where all the stars are CEMP-no. The star \Pis~ with its higher metallicity, is located at the limit between the lower and the higher A(C) band.


The oxygen abundance is also very high in \Pis: [O/Fe]=+2.25 (in normal EMP stars indeed, the oxygen abundance at this metallicity is close to [O/Fe]=+0.7). This strong enhancement is close to that observed in the galactic CEMP star CS\,22949-37: [O/Fe]=+1.98 \citep{DepagneHS02}.

\begin{figure}
\centering
\resizebox  {7.5cm}{8.5cm}
{\includegraphics {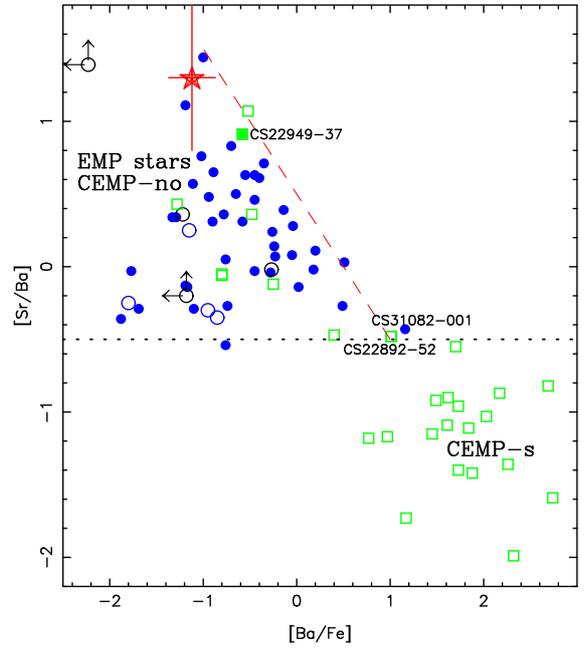} }
\caption[]{[Sr/Ba] vs. [Ba/Fe] in normal very metal-poor  stars in our Galaxy (blue filled circles),  in the stars in the dwarf galaxy Sculptor (black open circles), and in Tuc II (blue open circles). The green open squares represent a sample of Galactic CEMP stars studied from high-resolution spectra. The black dotted line at $\rm[Sr/Ba] \approx -0.5$ represents the pure r-process production as observed for example  in the EMP star CS\,31082-001 \citep{CayrelHB01,HillPC02} and the CEMP star CS\,22892-52 \citep{SnedenCI00,SnedenCL03}.\\
The CEMP stars with a high value of [Ba/Fe] and a low value of [Sr/Ba] have been enriched in neutron-capture elements by the ejecta of an AGB. Many CEMP stars have the same range of [Sr/Ba] as the normal EMP stars. They are CEMP-no. The red star  represents \Pis. It is very similar to the galactic C-rich giant CS\,22949-037 (filled green square). 
} 
\label{srba}
\end{figure}

\begin{figure}[ht]
\centering
\resizebox{\hsize}{!}
{\includegraphics {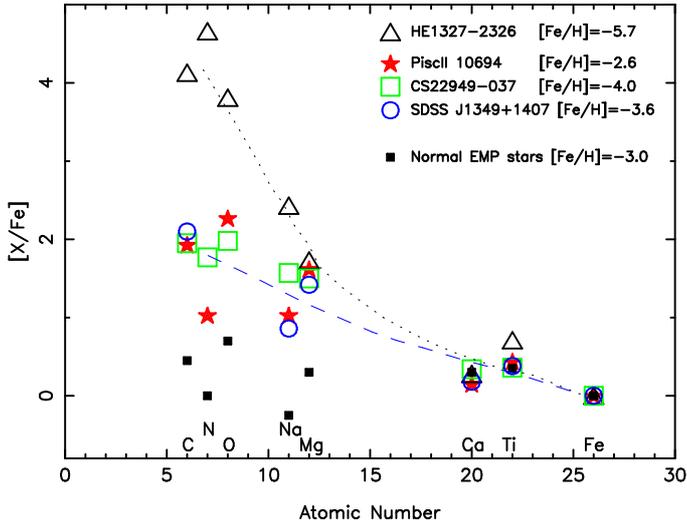} }
\caption[]{Abundance pattern of the elements in \Pis~ and three other galactic C-rich stars with [Fe/H]=-3.6, -4.0 and -5.7. The abundance pattern of the elements in \Pis~ is very similar to the pattern observed in CS\,22949-037 and SDSS\,J1349+1407.
In HE 1327-2326 which has an extremely low metal abundance, the enhancement of C, N, O, Na is stronger and the decrease steeper. In all the stars the abundance of the elements from Ca and heavier is normal. The dashed blue line shows the mean abundance pattern of the elements in \Pis, CS\,22949-037, and SDSS\,J1349+1407, and the dotted grey line the mean abundance pattern in HE 1327-2326.
} 
\label{pat}
\end{figure}

\subsubsection{Neutron-capture elements}
As said previously,  \Pis~ is not Ba-rich and thus it is a CEMP-no star.
Since we were able to obtain an estimation of the Sr abundance in this star (Table \ref{abund}), it is interesting to check whether it shares the same characteristics as the other CEMP-no stars. 

In Fig. \ref{srba} we have plotted [Sr/Ba] versus [Ba/Fe] for a sample of normal galactic metal-poor stars studied in the frame of the ESO Large Programme `First Stars'. All these stars are located in a well-defined region of this diagram \citep[see e.g.][]{SpiteSB18}.
Many Galactic CEMP stars are found in the same region as the normal EMP stars. 
They are CEMP-no with $\rm[Sr/Ba] > -0.5$. The limit $\rm[Sr/Ba]\approx -0.5$ corresponds to the production of a pure r-process  (dotted line on Fig.\ref{srba}).\\ 
On the other hand, in our Galaxy many CEMP stars are extremely rich in Ba ($\rm[Ba/Fe]>1$) and their ratio [Sr/Ba] is lower than the r-process limit. These stars have  likely been enriched by the ejecta of AGB stars or super-AGB stars forming neutron-capture elements through the s-process \citep[e.g.][]{BussoGW99} or the i-process \citep[e.g.][and references therein]{HampelSL16}.

\citet{JablonkaNM15} measure the abundance of Sr and Ba in stars of the dwarf galaxy Sculptor. In Fig. \ref{srba} we  compare the ratio [Sr/Ba] in the normal metal-poor stars in Sculptor and in our Galaxy; the stars of both galaxies occupy the same region of the diagram suggesting a similar origin for these elements. Unfortunately \citet{ChitiSF18} were  not able to measure the abundance of Ba and Sr in the C-rich star Sculptor\,11-1-4422, and thus we do not plot this star in  Fig. \ref{srba}.

Since the Pisces II stars are very faint, there is no determination of the abundance of the neutron-capture elements in these stars other than \Pis.
The position of \Pis~ in Fig. \ref{srba} suggests that it is a classical CEMP-no star, as is  CS\,22949-037  in our Galaxy. The position of \Pis~ in this diagram is also very close to the position of the normal very metal-poor giant HD\,122563, and it would be interesting to compare the neutron-capture element pattern in \Pis~and in HD122563. 
This will certainly be possible when a high-resolution spectrograph is available on the next generation of extremely large telescopes, but it may also be within reach of ESPRESSO \citep{espresso13,espresso14} fed by the four VLT telescopes. 

\subsection{Chemical imprint by the first stars}
The existence of a CEMP-no star in the UFD galaxy Pisces II (with $\rm L \sim 10^{4}L_{\odot}$) 
is consistent with theoretical predictions of cosmological models \citep[e.g.][] {SalvadoriST15} 
and hydro-dynamical simulations \citep[e.g.][]{JeonBB17} suggesting that these small galaxies 
might be the living relics of star-forming mini-halos hosting the first stars.
In these studies, long-lived CEMP-no stars form in gaseous environments predominantly imprinted        
by the chemical products of primordial low-energy faint supernovae (SNe) \citep[e.g.][] {BonifacioLC03,NomotoUM03}. 
The observed [C/Fe] = +1.23 and [Fe/H] = --2.6 of Pisces II 10694 suggest that its birth environment 
was likely polluted by both primordial faint SNe and normal Pop II stars exploding as core-collapse supernovae (\citealt{SalvadoriST15} section 4, and also \citealt{deBennassutiSS17}). 

\cite{SalvadoriST15} investigated the frequency of CEMP-no stars in dwarf galaxies with different luminosities. One of their key results is that the probability of observing CEMP-no stars, imprinted by primordial faint SNe, increases with decreasing galaxy luminosity and, on average, this probability is an order of magnitude higher in UFDs than in more massive classical Sculptor-like dwarf spheroidal galaxies. This is a direct consequence of the association between 
UFDs and low-mass mini-halos.
The detection of a single CEMP-no star in Pisces II do not allow us to make statistical comparisons with model predictions. However, our observational confirmation of one CEMP-no star out of the four 
stars identified by \cite{KirbySC15} at [Fe/H]$ < -2$ supports the idea that the probability of 
observing CEMP-no stars is high in ultra-faint dwarf galaxies.

\section{Conclusion}
In Pisces\,II only seven stars have been spectroscopically observed  \citep{KirbySC15}, and one of these stars \Pis, was found to be very metal-poor and was suspected to be C-rich.\\ 
--- We confirmed (Fig. \ref{CNO-LP}) that \Pis~ is a CEMP star. Its low lithium abundance (A(Li)<+0.2 dex) suggests that this cool giant star has undergone some extra mixing. As a consequence we adopted a correction of +0.7\,dex for [C/Fe] and --0.8dex for [N/Fe].\\
--- The ratios [N/Fe], [O/Fe], [Na/Fe], and [Mg/Fe] are also strongly enhanced in \Pis. The abundances of Ca to Ni are normal compared to the classical (not C-rich) EMP stars. The abundances of the $\alpha$ elements Mg, Ca, and Ti are only slightly enhanced relative to Fe, as  is generally observed in metal-poor stars.\\
--- The abundance pattern in \Pis~ (Fig. \ref{pat}) is similar to that observed in the galactic C-rich giant  CS\,22949-037 \citep{DepagneHS02}, and in SDSS\,J1349+1407, a turn-off CEMP star recently reported by \citet{BonifacioCS18}. These three stars have a metallicity  [Fe/H] between --2.6 and --4.0.\\ 
In Fig. \ref{pat} we have also added the abundance pattern of the galactic extremely metal-poor C-rich subgiant HE1327-2326: $\rm[Fe/H]=-5.7$ following \citet{FrebelCE08}. 
In this star, the enhancement of C,N,O, and Na is much stronger, but the decrease of this enhancement with the atomic number is steeper. All the stars in Fig. \ref{pat} have a normal [Ca/Fe] ratio.\\
--- The carbon abundance in \Pis~ (A(C)=7.73 with the extra-mixing correction) put this star at the limit between the lower and the higher A(C) band  \citep[Fig. \ref{Cab} and ][]{SpiteCB13,BonifacioCS15}.\\ 
--- The observed chemical properties of \Pis~ suggest that it is a CEMP-no star.
The abundance of strontium and barium are very low in \Pis, as they are in the classical Galactic CEMP-no star CS\,22949-037. 
Our observational confirmation of a CEMP-no star in the UFD galaxy Pisces II agrees with the 
predictions of theoretical models that follow early cosmic star formation in low-mass mini-halos and 
trace the chemical signature of zero-metallicity stars exploding as faint SN (e.g. Salvadori et al. 2015). 
Hence, our results suggest that the UFD galaxy Pisces II might be the living fossil of a star-forming
mini-halo hosting the first stars.
To make statistical comparisons between model predictions and observations of UFD galaxies, larger 
samples of CEMP-no stars are required. Our observational findings show that the Pisces II galaxy is a 
perfect place to look for more iron-poor, highly C-enhanced CEMP-no stars similar to those 
found in the Galactic halo and in the faintest UFD Segue I.

\begin {acknowledgements} 
This work has made use of data from the European Space Agency (ESA) mission
{\it Gaia} (\url{https://www.cosmos.esa.int/gaia}), processed by the {\it Gaia}
Data Processing and Analysis Consortium (DPAC,
\url{https://www.cosmos.esa.int/web/gaia/dpac/consortium}). Funding for the DPAC
has been provided by national institutions, in particular the institutions
participating in the {\it Gaia} Multilateral Agreement.
Stefania Salvadori is supported by the Italian Ministry of Education, University, and Research
(MIUR) through a Rita Levi Montalcini Fellowship.
\end {acknowledgements}

\bibliographystyle{aa}
{}

\appendix
\section{}
\begin{table*}[h]
\begin{center}   
\caption[]{ 
 LTE determination of the abundance $\rm A(X) = log(N_{X} / N_{H}) + 12$ of the elements in \Pis.}
\label{lines}
\begin{tabular}{ccccrrcccccrrrrrrr}
\hline
Element & wavelength &\kiex & \loggf & A(X)&~$\vert$~~&Element & wavelength &\kiex & \loggf  & A(X) \\
        & in \AA     &      &        &     &~$\vert$~~&        &  in \AA  \\
\hline
C\,(CH) & 4260-4300  &  -  & - & 6.68     &~$\vert$~~&\Cad    & 8498.1  & 1.69   & -1.429 & 4.46  \\
C\,(CH) & 4294-4316  &  -  & - & 6.74     &~$\vert$~~&\Cad    & 8542.1  & 1.70   & -0.463 & 4.11  \\
C\,(CH) & 4320-4340  &  -  & - & 7.28     &~$\vert$~~&\Cad    & 8662.2  & 1.69   & -0.723 & 4.15  \\
C\,(CH) & 4360-4380  &  -  & - & 7.23     &~$\vert$~~&        &        &        &        &         \\
        &            &     &   &          &~$\vert$~~&\Tiu    & 6258.1  & 1.44   & -0.390 &  2.74 \\
C\,(\Cd)& 5030-5140  &  -  & - & 7.24     &~$\vert$~~&\Tiu    & 6258.7  & 1.46   & -0.280 &  2.74 \\
C\,(\Cd)& 5120-5165  &  -  & - & 7.31     &~$\vert$~~&        &        &        &        &         \\
      &              &     &   &          &~$\vert$~~&\Feu    & 6065.5  & 2.61   & -1.530 &  5.04 \\
N\,(CN) & 4120-4200  &  -  & - & 6.73     &~$\vert$~~&\Feu    & 6136.6  & 2.45   & -1.400 &  4.94 \\
N\,(CN) & 4200-4215  &  -  & - & 6.88     &~$\vert$~~&\Feu    & 6137.7  & 2.59   & -1.403 &  4.94 \\
N\,(CN) & 8070-8350  &  -  & - & 7.35     &~$\vert$~~&\Feu    & 6191.6  & 2.43   & -1.417 &  4.88 \\
N\,(CN) & 8308-8350  &  -  & - & 7.09     &~$\vert$~~&\Feu    & 6213.4  & 2.22   & -2.482 &  5.00 \\
N\,(CN) & 8362-8388  &  -  & - & 7.06     &~$\vert$~~&\Feu    & 6219.3  & 2.20   & -2.433 &  4.94 \\
N\,(CN)K*& 8308-8350 &  -  & - & 7.15     &~$\vert$~~&\Feu    & 6230.7  & 2.56   & -1.281 &  4.89 \\
N\,(CN)K*& 8362-8388 &  -  & - & 7.36     &~$\vert$~~&\Feu    & 6252.6  & 2.40   & -1.687 &  4.97 \\
        &         &        &        &     &~$\vert$~~&\Feu    & 6256.4  & 2.45   & -2.408 &  4.97 \\
$\rm [\Ou]$& 6300.3  & 0.00&  -9.78 & 8.30&~$\vert$~~&\Feu    & 6265.1  & 2.18   & -2.550 &  4.82 \\
$\rm [\Ou]$& 6363.8  & 0.00&-10.258 & 8.55&~$\vert$~~&\Feu    & 6393.6  & 2.43   & -1.432 &  4.83 \\
        &         &        &        &     &~$\vert$~~&\Feu    & 6400.0  & 3.60   & -0.290 &  4.70 \\
\Nau    & 5890.0  & 0.00   & -0.194 & 4.73&~$\vert$~~&\Feu    & 6421.4  & 2.28   & -2.027 &  4.84 \\
        &         &        &        &     &~$\vert$~~&\Feu    & 6430.9  & 2.18   & -2.006 &  4.86 \\
\Mgu    & 4703.0  & 4.35   & -0.440 & 6.32&~$\vert$~~&\Feu    & 6678.0  & 2.69   & -1.418 &  5.13 \\
\Mgu    & 5167.3  & 2.71   & -0.931 & 6.68&~$\vert$~~&        &        &        &        &         \\
\Mgu    & 5172.7  & 2.71   & -0.450 & 6.78&~$\vert$~~&\Niu    & 6643.6  & 1.68   & -2.300 &  3.99 \\
\Mgu    & 5183.6  & 2.72   & -0.239 & 6.90&~$\vert$~~&\Niu    & 6767.8  & 1.83   & -2.170 &  3.67 \\
\Mgu    & 5528.4  & 4.35   & -0.498 & 6.11&~$\vert$~~&        &        &        &        &         \\
\Mgu    & 8717.8  & 5.93   & -0.865 & 6.70&~$\vert$~~&\Srd    & 4077.7  & 0.00   & +0.167 &    0.20:\\
\Mgu    & 8736.0  & 5.95   & -0.350 & 6.44&~$\vert$~~&\Srd    &10036.7  & 1.81   & -1.312 &  0.80:\\
        &         &        &        &     &~$\vert$~~&        &        &        &        &         \\
\Cau    & 5598.5  & 2.52   & -0.087 & 3.76&~$\vert$~~&\Bad    & 4554.0  & 0.00   & +0.170 & -2.15:\\
\Cau    & 5857.4  & 2.93   & +0.240 & 4.23&~$\vert$~~&\Bad    & 4934.1  & 0.00   & -0.150 & -1.66 \\
\Cau    & 6102.7  & 1.88   & -0.793 & 4.06&~$\vert$~~&\Bad    & 6141.7  & 0.70   & -0.076 & -1.58 \\
\Cau    & 6122.2  & 1.89   & -0.316 & 3.56&~$\vert$~~&\Bad    & 6496.9  & 0.60   & -0.377 & -1.31 \\
\Cau    & 6162.2  & 1.90   & -0.090 & 3.90&~$\vert$~~&\\
\Cau    & 6439.1  & 2.53   & +0.390 & 3.70&~$\vert$~~&\\
\Cau    & 6493.8  & 2.52   & -0.109 & 3.92&~$\vert$~~&\\
\hline    
\multicolumn{5}{l}{* Abundance measured on the Kirby spectrum.}
\end{tabular}  
\end{center}   
\end{table*}

\end{document}